\documentclass[3p,times,procedia]{elsarticle}
\usepackage{nupha_ecrc}

\volume{00}
\firstpage{1}
\journalname{Nuclear Physics A}
\runauth{Subhash Singha (for STAR Collaboration)}
\jid{nupha}
\jnltitlelogo{Nuclear Physics A}

\usepackage{amssymb}

\usepackage[figuresright]{rotating}
\usepackage{natbib}

\begin{document}

\begin{frontmatter}
\dochead{XXVIIIth International Conference on Ultrarelativistic Nucleus-Nucleus Collisions\\ (Quark Matter 2019)}

\title{Measurement of global spin alignment of $K^{*0}$and $\phi$ vector mesons using the STAR detector at RHIC}
\author{Subhash Singha (for the STAR Collaboration)}
\address{Institute of Modern Physics Chinese Academy of Sciences,
  Lanzhou, Gansu, China 73000 \\ subhash@impcas.ac.cn}
 
\begin{abstract}
We report the transverse momentum ($p_{\mathrm{T}}$) and
centrality dependence of global spin alignment ($\rho_{00}$) of $K^{*0}$ vector
meson at midrapidity ($|y|<0.5$) in Au + Au collisions at $\sqrt{s_{NN}}$ =
54.4 and 200~GeV with the STAR experiment at RHIC. The $K^{*0}$ results are compared to that of $\phi$ meson. At low-$p_{\mathrm{T}}$ region and midcentral collisions,
the $K^{*0}$ $\rho_{00}$ is found to be smaller than 1/3 with about 4$\sigma$ significance,
while that of $\phi$ meson is observed to be larger than
1/3 with about 3$\sigma$ significance. The $\rho_{00}$ results are compared
between RHIC and  LHC energies. The physics implication of our results
is also discussed.
\end{abstract}


\end{frontmatter}

\section{Introduction}
\label{proc-intro}

In non-central heavy-ion collisions, a large initial global angular
momentum ($\sim 10^{4} \hslash$ ) is
expected~\cite{becattini}. This can induce a non-vanishing polarization
for hadrons with non-zero spin via spin-orbit coupling. The measurement of spin polarization can offer new insight into the initial conditions and dynamics of the Quark-Gluon
Plasma (QGP)~\cite{liang0,betz}. The STAR Collaboration reported significant non-zero
$\Lambda$ polarization at RHIC energies~\cite{star_lambda_nature,star_lambda_prc}.
This provides the first experimental evidence of the vorticity
of the QGP medium induced by the initial angular momentum. The spin
alignment of vector mesons can also be used to
probe the vorticity~\cite{liang1}. The vector meson global spin alignment
is quantified by the diagonal element of the spin density matrix ($\rho_{00}$)~\cite{schiling}. It
is measured from the angular distribution of the decay daughter of the vector meson:
\begin{equation}
\label{eqn1}
\frac{dN}{d \rm cos \theta^{*}} \propto  \big[ (1-\rho_{00}) +
(3\rho_{00}-1) \rm cos^{2}  \theta^{*}  \big],
\end{equation}
where $\theta^{*}$ is the angle between the polarization axis and momentum direction of the
daughter particle in the rest frame of parent particle. For global
spin alignment, the polarization axis is chosen as the direction
perpendicular to the reaction plane which is correlated with the direction of the 
angular momentum of the colliding system. In the absence of
spin alignment, the value of $\rho_{00}$ is expected to be 1/3. Any
deviation of $\rho_{00}$ from 1/3 indicate a net spin
alignment of vector mesons. Recent model calculations indicated that
the $\rho_{00}$ due to vorticity of the medium is expected to be smaller
than 1/3, while that induced by the initial magnetic field can be
larger or smaller than 1/3 depending on the electric charge of the
vector meson~\cite{yang}. It is also been predicted that different hadronization scenarios, such as the
fragmentation and coalescence mechanisms can cause $\rho_{00}$ to be
larger and smaller than 1/3 ~\cite{liang1}. The vector mesons,
$K^{*0}$ and $\phi$, are expected to be produced predominantly from
primordial production, unlike hyperons which are expected to have
large resonance decay contribution. Another advantage is that the spin
alignment of vector mesons are generally additive, whereas hyperons
polarization are subject to local cancellation effects. Moreover, the lifetime of these
vector mesons differ by a factor of ten, so they can carry informations
of the medium from different time scale during its evolution.

\section{Analysis method}
\label{proc-meth}
This proceedings report the measurement of $K^{*0}$ $\rho_{00}$ at midrapidity ($|y|<0.5$) 
in Au+Au  collisions at $\sqrt{s_{\rm NN}}$ = 54.4 and 200~GeV. The minimum-bias
events are selected by coincidence of east and west Vertex Position Detectors~\cite{vpd}. The charge
particle tracking is performed using the Time Projection Chamber
(TPC)~\cite{tpc}. The collision centrality is determined from the number of
charged particles within $|\eta| <$ 0.5 and corrected for triggering
efficiency using a Monte Carlo Glauber
simulation~\cite{centrality_glauber}. The $2^{nd}$-order event plane
(experimental approximate of reaction plane) is
reconstructed using tracks inside the TPC. The particle identification is done using
the specific ionization energy loss in TPC gas volume and the velocity of
particles ($1/\beta$) measured by the Time-of-Flight (TOF) detector~\cite{tof}. The $K^{*0}$ ($\overline{K^{*0}}$) is
reconstructed via hadronic decay channel: $K^{*0} (\overline{K^{*0}}) \rightarrow K^{+}
\pi^{-}$ ($K^{-} \pi^{+}$) (branching ratio:
66$\%$)~\cite{pdg}. Measurement of $K^{*0}$ and $\overline{K^{*0}}$
are averaged and they are collectively referred to as $K^{*0}$.
The combinatorial background is estimated from a
pair rotation technique. The invariant mass signal is obtained after the
subtraction of the combinatorial background. The
signal is fitted with a Breit-Wigner distribution and a second-order
polynomial function to take care of residual background. 
The yield is estimated by integrating signal histogram bins within the range:
($m_{0}-3\Gamma, m_{0}+3\Gamma$), where $m_{0}$ and
$\Gamma$ are the invariant mass peak position and width of $K^{*0}$.
The $K^{*0}$ yield is obtained in five $\rm cos \theta^{*}$ bins,
where the $\theta^{*}$ is the angle between the direction
perpendicular to the $2^{nd}$-order event plane and the momentum direction
of daughter kaon in the rest frame of $K^{*0}$. The yield in each $\rm cos \theta^{*}$ bin is
then corrected for detector acceptance and efficiency using a Monte
Carlo embedding. We extract the observed $\rho_{00}$ (denoted as $\rho_{00}^{\rm obs}$) by fitting
the yield vs. $\rm cos \theta^{*}$ distribution using
Eq.~\ref{eqn1}. The $\rho_{00}^{\rm obs}$ is then corrected for event plane
resolution, following method detailed in~\cite{aihong}, to obtain $\rho_{00}$:
\begin{equation}
\label{eqn2}
\rho_{00} - \frac{1}{3} = \frac{4}{1+3R} (\rho_{00}^{obs} - \frac{1}{3}),
\end{equation}
where $R$ is the TPC $2^{nd}$-order event plane resolution, estimated from the
correlation of two sub-events~\cite{flow}. 
\section{Results and discussions}
\label{proc-result}

\begin{figure}[!htb]
\begin{center}
\includegraphics[scale=0.37]{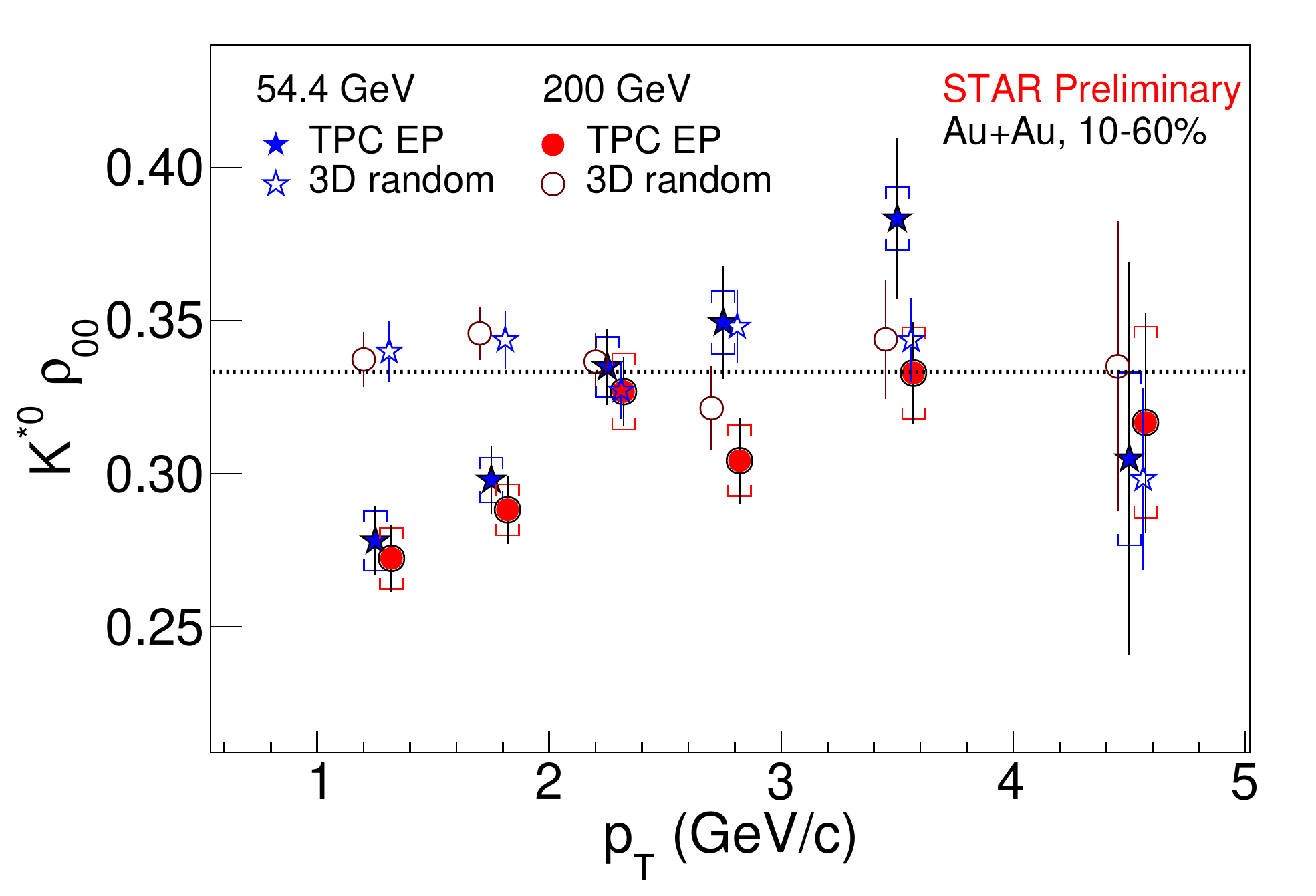}
\includegraphics[scale=0.37]{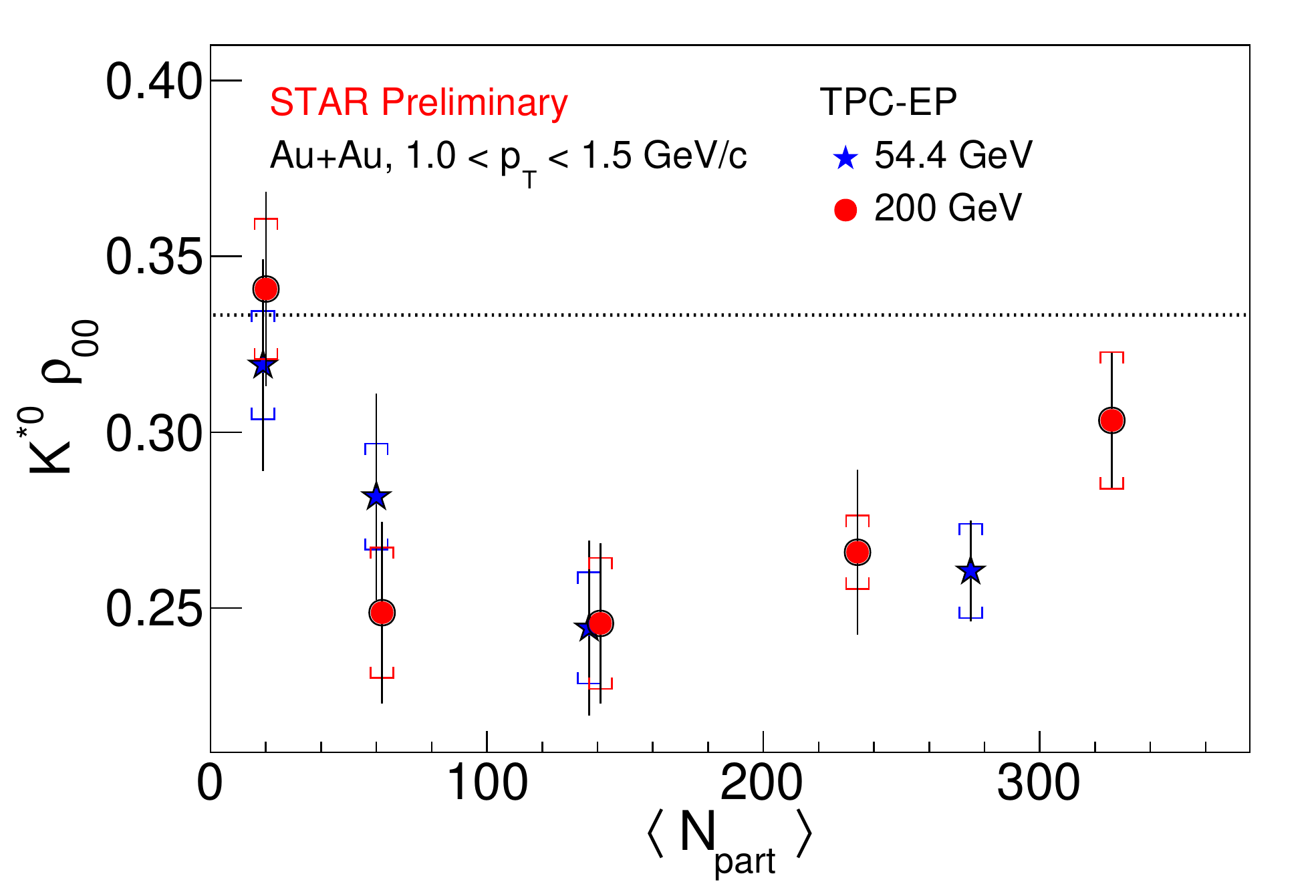}
\vspace*{-3mm}
\caption{ Left panel: The solid circles and star markers
  present the $K^{*0}$ $\rho_{00}$ using the 2$^{nd}$-order event plane as function of $p_{\mathrm{T}}$
for 10-60$\%$ central Au+Au collisions at $\sqrt{s_{\rm NN}}$ = 54.4
and 200~GeV, respectively. The open markers present the same using a three-dimensional (3D)  random plane. Right panel: The solid circles and star markers
  present the $K^{*0}$ $\rho_{00}$ using the 2$^{nd}$-order event plane as function of $ \langle N_{\rm part} \rangle$ for $1.0 < p_{\mathrm{T}} < 1.5$ GeV/c. The vertical bars and caps denote
  statistical and systematic uncertainties, respectively. 
}
\label{fig1}
\end{center}
\end{figure}
\vspace*{-3mm}
\subsection{Transverse momentum ($p_{\mathrm{T}}$) dependence}
The solid star and circle markers in the left panel of Fig.~\ref{fig1} present the $K^{*0}$
$\rho_{00}$ measured using the 2$^{nd}$-order event plane
as a function of $p_{\mathrm{T}}$
for 10-60$\%$ central Au+Au collisions at $\sqrt{s_{\rm NN}}$ = 54.4
and 200~GeV. The open markers present the $\rho_{00}$ with respect to a
three-dimensional (3D) random plane which is not expected to be correlated with the angular
momentum direction. We observe that the $\rho_{00}$ for $p_{\mathrm{T}} < $ 2.0 GeV/c is smaller than 1/3 with
about 4$\sigma$ significance, while for higher $p_{\mathrm{T}}$ region the
$\rho_{00}$ is consistent with 1/3 within uncertainties. The $\rho_{00}$
with respect to the 3D random plane is found to be consistent with 1/3
as expected. The observed deviation of $K^{*0}$ $\rho_{00}$ in low-$p_{\mathrm{T}}$ region can be qualitatively explained by models that consider
the hadronization of polarized quarks via coalescence mechanism~\cite{liang1}. But to
date, there is no quantitative estimate of $K^{*0}$ $\rho_{00}$
available from such models. The $\rho_{00}$ of $\phi$
meson for $p_{\mathrm{T}}$=1.0--2.0~GeV/c in midcentral Au+Au collisions at
$\sqrt{s_{\rm NN}}$ =  200~GeV (presented in QM2018)~\cite{zhou} is
observed to be larger than 1/3. The $\phi$ $\rho_{00}$ measurement
does not fit into the naive quark coalescence or fragmentation model. 

\subsection{Centrality ($\langle  N_{\rm part} \rangle $) dependence}
The right panel in Fig.~\ref{fig1} shows the $K^{*0}$ $\rho_{00}$
as function of average number of participating nucleons ($\langle  N_{\rm part} \rangle $)
for  $1.0 < p_{\mathrm{T}} < 1.5$ GeV/c in Au+Au collisions at
$\sqrt{s_{\rm NN}}$ = 54.4 and 200~GeV. We observed a clear centrality dependence with
the maximum deviation of $\rho_{00}$ from 1/3 in midcentral
collisions. For peripheral collisions, the $K^{*0}$ $\rho_{00}$ is
consistent with 1/3 while for most central collisions it is close to 1/3. 
The $\rho_{00}$ of $\phi$ mesons (presented in QM2018)~\cite{zhou} shows similar centrality
 dependence, but with an opposite trend and the $\rho_{00}$ is larger than
 1/3 in midcentral collisions. Current models can not simultaneously
 explain the observed centrality dependence of $\rho_{00}$ of $K^{*0}$  and $\phi$ mesons.

\begin{figure}[!htb]
\begin{center}
\includegraphics[scale=0.37]{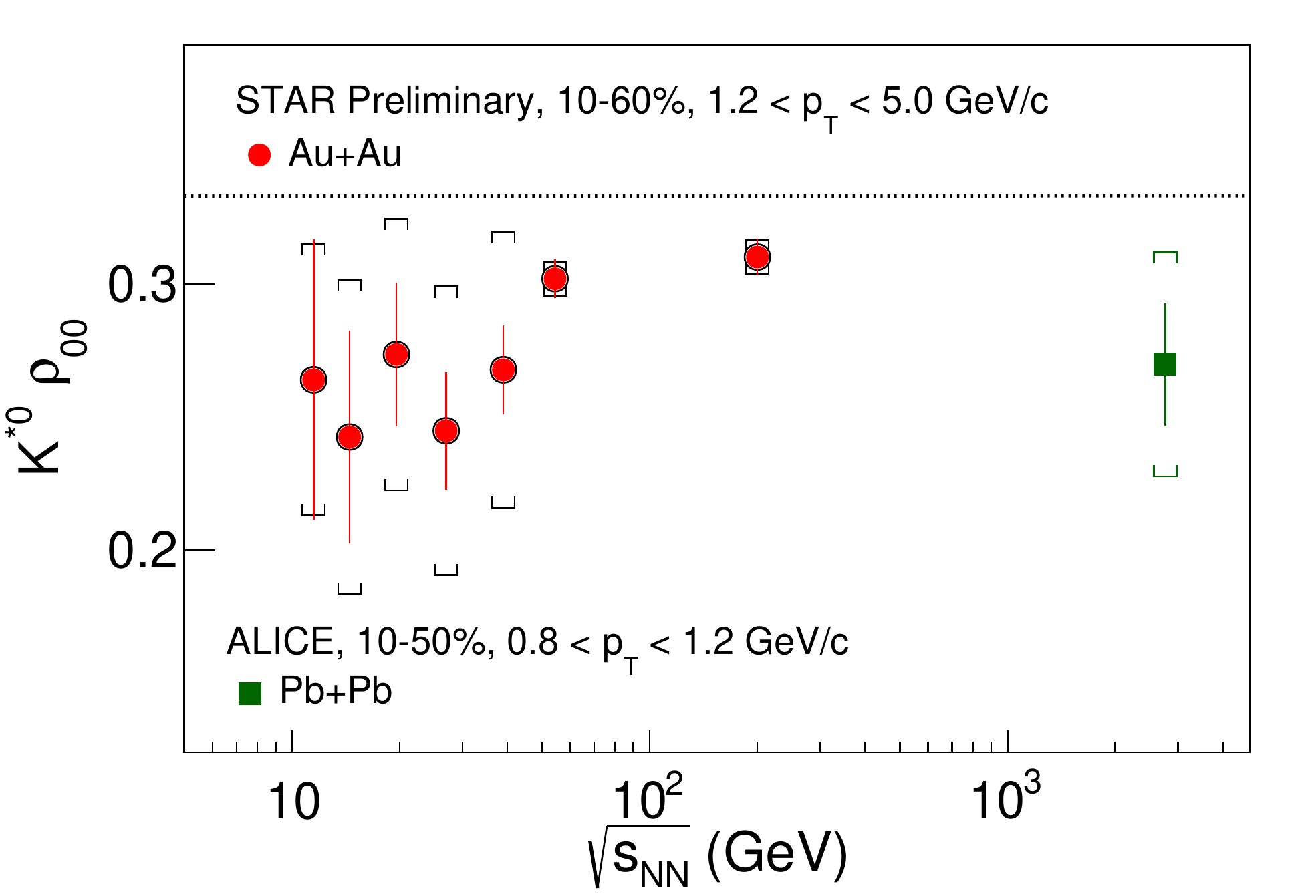}
\includegraphics[scale=0.37]{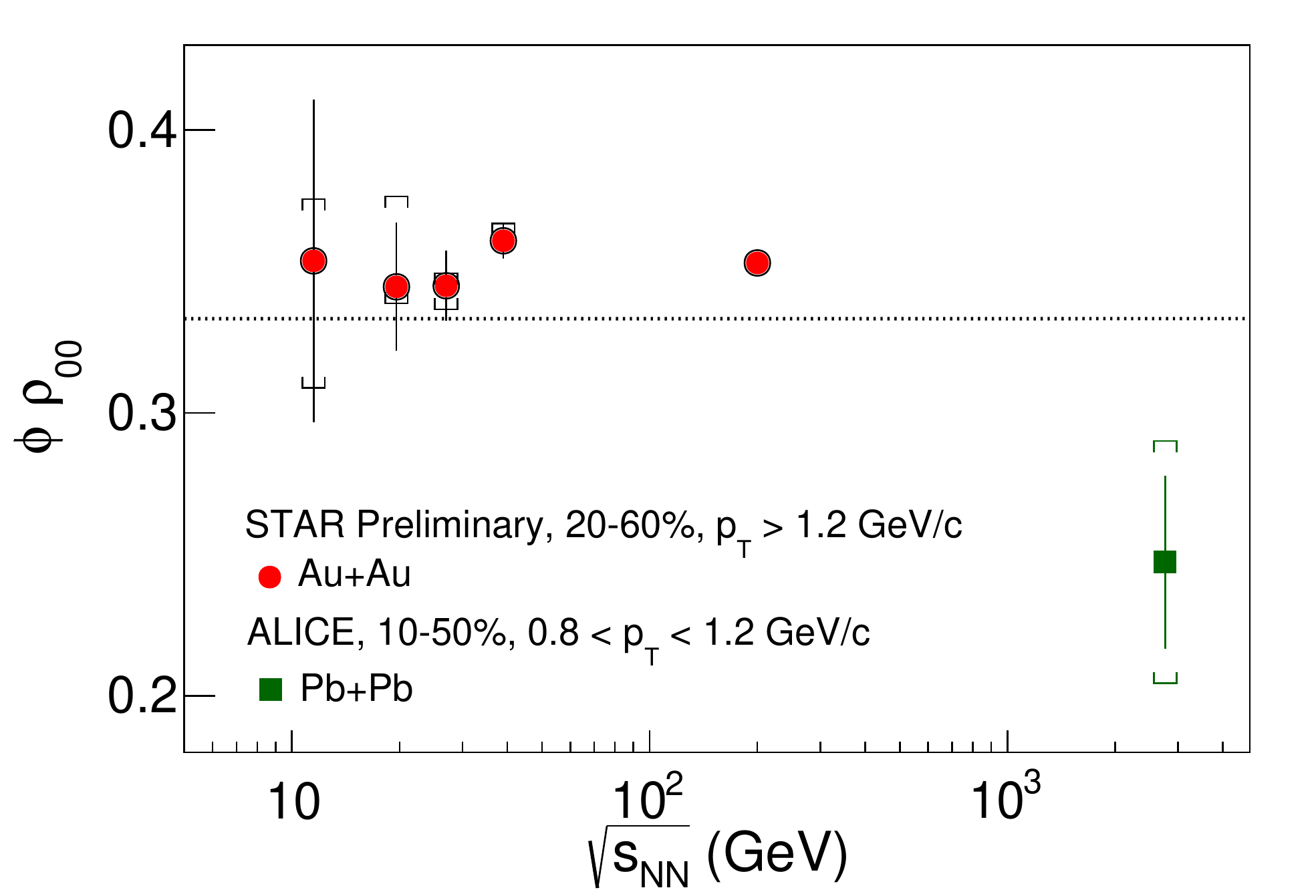}
\vspace*{-3mm}
\caption{ Left panel: beam-energy dependence of $K^{*0}$
  $\rho_{00}$ in midcentral collisions. Right panel: beam-energy dependence of $\phi$
  $\rho_{00}$ in midcentral collisions. In both panels, the vertical
  bars and caps denote statistical and systematic uncertainties,
  respectively. 
}
\label{fig2}
\end{center}
\end{figure}
\vspace*{-2mm}

\subsection{Beam-energy ($\sqrt{s_{\rm NN}}$) dependence}
The left panel in Fig.~\ref{fig2} presents the beam-energy dependence
of $K^{*0}$ $\rho_{00}$ in midcentral collisions. The new measurements
for Au+Au collisions at $\sqrt{s_{\rm NN}}$ = 54.4 and 200~GeV are
compared to those from Au+Au collisions at  $\sqrt{s_{\rm NN}}$ =
11.5 -- 39 GeV reported in~\cite{zhou} and from Pb+Pb collisions at $\sqrt{s_{\rm NN}}$ =
2.76~TeV~\cite{alice}. The $K^{*0}$ $\rho_{00}$ for low $p_{\mathrm{T}}$ and midcentral
collisions is found to be smaller than 1/3 and within the present
uncertainties no beam-energy dependence is observed. The right
panel in Fig.~\ref{fig2} presents the beam energy dependence of $\phi$
$\rho_{00}$. The STAR results~\cite{zhou} from Au+Au collisions at $\sqrt{s_{\rm
    NN}}$ =11.5--200~GeV are compared to the measurements from LHC
energies~\cite{alice}. While the $\phi$ $\rho_{00}$ in midcentral
collisions at RHIC energies is observed to be larger than 1/3 (about ~3$\sigma$ significance at 39 and 200
GeV), it is found to be smaller than 1/3 at the LHC energy (about
~2$\sigma$ significance). The trend of $\phi$ $\rho_{00}$ at RHIC
energies can be explained by a recent model calculation that considers the
existence of coherent mesonic field~\cite{phi-field}. Note that
the calculation mentioned above does not exist for the $K^{*0}$ meson.

\section{Summary and conclusion}
\label{proc-summary}

We presented $p_{\mathrm{T}}$ and centrality dependence of $\rho_{00}$ of
$K^{*0}$ meson for Au+Au collisions at $\sqrt{s_{NN}}$= 54.4 and 200
GeV. At low $p_{\mathrm{T}}$ and midcentral collisions, the $K^{*0}$
$\rho_{00}$ is observed to be smaller than 1/3 with 4$\sigma$ significance for both beam
energies. This is an indication of $K^{*0}$ spin alignment for both beam energies.
For midcentral collisions, while the $K^{*0}$ $\rho_{00}$ is found
to be smaller than 1/3, the $\phi$ $\rho_{00}$ is observed to be larger
than 1/3. It could be due to the different lifetime of these vector
mesons and different responses to the vorticity of the medium at different time scales. No
current theoretical model can explain simultaneously  the trend of $K^{*0}$ and
$\phi$ $\rho_{00}$. Within the current precision, no significant
beam-energy dependence is observed for $K^{*0}$ $\rho_{00}$. The data from
the $2^{nd}$ phase of the Beam Energy Scan in RHIC will improve the precision of
the low energy data. The $p_{\mathrm{T}}$ and centrality dependence of
$\rho_{00}$ of $K^{*0}$ is qualitatively similar between RHIC and LHC energies.

From the current theoretical understanding, the global hyperon polarization
($P_{H}$) is proportional to the quark polarization ($P_{q}$): $P_{H} \propto
P_{q}$, while the spin alignment, $\rho_{00} \propto P_{q}^{2}$. Based
on the above assumptions and the input of $P_{q}$ from $\Lambda$
polarization measurement, the expected $\rho_{00}$ is close to 1/3.
Hence, the current large deviation of
$\rho_{00}$ is surprising and poses challenges to theoretical
understanding. Given the $\rho_{00}$ can depend on multiple physics
mechanisms, e.g. the vorticity, magnetic
field, hadronization scenarios and mesonic fields, more
theoretical efforts are required for understanding of the
data. 

\begin {table}
\begin{center}
\caption{Sumary of $\rho_{00}$ and $P_{H}$ measurements at RHIC and LHC} \label{tab-summ}
\begin{tabular}{ |l|c|c|c|c| } 
 \hline
 Species & Quark content & $J^{P}$& $\rho_{00}/P_{H}$ at top-RHIC & $\rho_{00}/P_{H}$ at LHC\\ 
 \hline
 $K^{*0}$& $d\bar{s}$ & $1^{-}$ & $\rho_{00} < 1/3\; (\sim 4\sigma)$& $\rho_{00} < 1/3\; (\sim 3\sigma)$\\ 
 \hline
 $\phi$& $s\bar{s}$ & $1^{-}$ & $\rho_{00} > 1/3\; (\sim 3\sigma)$& $\rho_{00} < 1/3\; (\sim 2\sigma)$\\ 
 \hline
 $\Lambda$& $uds$ & $1/2^{+}$ & $ P_{H} > 0 ; (\sim 4\sigma)$& $P_{H} \sim 0\; (\sim 1\sigma)$\\ 
\hline
\end{tabular}
\end{center}
\end {table}


\bibliographystyle{elsarticle-num}
\bibliography{<your-bib-database>}

\begin{thebibliography}{00}
\bibitem{becattini} F. Becattini, F. Piccinini, J. Rizzo,
  Phys. Rev. {\bf C77} (2008) 024906.
\bibitem{liang0} Z.-T. Liang, X.-N. Wang, Phys. Rev. Lett. {\bf 94} (2005) 102301, [Erratum: Phys. Rev. Lett.96,039901(2006)].
\bibitem{betz} B. Betz, M. Gyulassy, G. Torrieri, Phys. Rev. {\bf C76} (2007) 044901.
\bibitem{star_lambda_nature} L. Adamczyk, {\it et al.}, [STAR Collaboration], Nature {\bf 548} (2017) 62–65
\bibitem{star_lambda_prc} J. Adam, {\it et al.}, [STAR Collaboration], Phys. Rev. {\bf C98} (2018) 014910.
\bibitem{liang1} Z.-T. Liang, X.-N. Wang, Phys. Lett. {\bf B629} (2005) 20–26. 
\bibitem{schiling} K. Schilling, P. Seyboth, G. E. Wolf, Nucl. Phys. {\bf B15} (1970) 397–412, [Erratum: Nucl. Phys.B18,332(1970)]. 
\bibitem{yang} Y.-G. Yang, R.-H. Fang, Q. Wang, X.-N. Wang {\bf C97} (3) (2018) 034917
\bibitem{vpd} W. J. Llope, {\it et al.}, [STAR Collaboration], Nucl. Instrum. Meth. {\bf A522} (2004) 252–273.
\bibitem{tpc} M. Anderson, {\it et al.}, [STAR Collaboration], Nucl. Instrum. Meth. {\bf A499} (2003) 659–678.
\bibitem{centrality_glauber} B. I. Abelev, {\it et al.}, [STAR Collaboration], Phys. Rev. {\bf C79} (2009) 034909.
\bibitem{tof} B. Bonner {\it et al.}, Nucl. Instrum. Meth. {\bf A508} (2003) 181–184.
\bibitem{pdg} M. Tanabashi, {\it et al.},, Review of Particle Physics, Phys. Rev. {\bf D98} (3) (2018) 030001.
\bibitem{aihong} A. H. Tang, B. Tu, C. S. Zhou, Phys. Rev. {\bf C98} (4) (2018) 044907
\bibitem{flow} A. M. Poskanzer, S. A. Voloshin, Phys. Rev. {\bf C58} (1998) 1671–1678.
\bibitem{zhou} C. Zhou, [for STAR Collaboration], Nucl. Phys. {\bf A982} (2019) 559–562. 
\bibitem{alice} S. Acharya, {\it et al.}, [ALICE Collaboration],
  arXiv:1910.14408.
\bibitem{phi-field} X. Sheng, L. Oliva, Q. Wang, arXiv:1910.13684 
\end{thebibliography}

\end{document}